\documentclass[11pt]{article}

\usepackage{amsmath,amssymb}
\pdfoutput=1

\usepackage{epsfig}
\usepackage{graphicx}               
\usepackage{url}
\usepackage{hyperref}
\usepackage{cite}

\newcommand{\nc}{\newcommand}

\nc{\beq}{\begin{equation}}
\nc{\eeq}{\end{equation}}
\nc{\beqa}{\begin{eqnarray}}
\nc{\eeqa}{\end{eqnarray}}
\nc{\bea}{\begin{eqnarray}}
\nc{\eea}{\end{eqnarray}}
\nc{\ra}{\rightarrow}
\nc{\lsim}{\begin{array}{c}\,\sim\vspace{-21pt}\\< \end{array}}
\nc{\gsim}{\begin{array}{c}\sim\vspace{-21pt}\\> \end{array}}
\nc{\Tr}{{\rm Tr}}
\nc{\slsh}{\slash\hspace*{-0.22cm}}

\def\be{\begin{equation}}
\def\ee{\end{equation}}
\def\bea{\begin{eqnarray}}
\def\eea{\end{eqnarray}}
\def\bit{\begin{itemize}}
\def\eit{\end{itemize}}

\newcommand{\gev}{{\rm GeV}}

\def\jet{J_{E_T}}

\title{
\vspace*{-2.3cm}
\begin{flushright}
\normalsize{
  }
\end{flushright}
\vspace{1.5cm}
\Large
\boldmath{$J_{E_T}$}:
\bf{A Global Jet Finding Algorithm}
\vspace*{1.0cm}
}

\author{
Yang Bai,$^{a}$ Zhenyu Han$^{b}$ and Ran Lu$^{a}$
\vspace{5mm}
\\
$^{a}$ \normalsize\emph{Department of Physics, University of Wisconsin, Madison, WI 53706, USA}  \vspace{1mm} \\
$^{a}$ \normalsize\emph{Institute for Theoretical Science, University of Oregon, Eugene, OR 97403, USA}
}

\date{}

\begin{document}
\setcounter{page}{0}
\maketitle

\vspace*{1cm}
\begin{abstract}
 \normalsize{ 
We introduce a new jet-finding algorithm for a hadron collider based on maximizing a $J_{E_T}$ function for all possible combinations of particles in an event. This function prefers a larger value of the jet transverse energy and a smaller value of the jet mass. The jet shape is proved to be a circular cone in Cartesian coordinates with the geometric center shifted from the jet momentum toward the central region. The jet cone size shrinks for a more forward jet. We have implemented our $J_{E_T}$ algorithm with a reasonable running time scaling as $N n^3$, where ``$N$'' is the total number of particles and ``$n$'' $(\ll N)$ is the number of particles in a fiducial region. Many features of our $J_{E_T}$ jets are similar to anti-$k_t$ jets, including the reconstructed jet momentum and the  ``back-reaction" from soft contamination. Nevertheless, when the jet parameters in the two algorithms are matched using QCD jets, we find that the $J_{E_T}$ algorithm has a larger efficiency than anti-$k_t$ for identifying objects with hard splittings such as a $W$-jet. 
 }
\end{abstract}

\thispagestyle{empty}
\newpage

\setcounter{page}{1}

\baselineskip18pt

\vspace{-3cm}

\section{Introduction}
\label{sec:intro}
In high energy physics, the concept of a ``jet'' originates from the fact that an energetic QCD parton, either a quark or a gluon, cannot be directly observed in a collision. Instead, a bundle of particles are produced from the hard parton, which together form the jet that can be detected in a collider detector. Then the momentum of the parton can be obtained by measuring the momentum of the jet it initiates. Recently, the notion of jet has been extended to include highly boosted massive particles that decay to QCD partons. Since the decay products of such particles are collimated, they behave as a single jet using traditional jet-finding methods. In the search for new physics, jets with different origins could become backgrounds of one another. Therefore, it is essential to study jet substructure to identify jets with different quantum numbers or origins. Recently, we have seen increased activities of jet substructure studies including $W$/Higgs/top jet tagging, quark-gluon discrimination and jet charge measurement (see Ref.~\cite{jetography, boost2011, boost2012, Shelton:2013an} for reviews and references therein).

In order to study jet properties including its momentum and substructure, the first step is to reconstruct the jet using a jet-finding algorithm. The majority of existing jet algorithms fall into two categories: cone algorithms \cite{Sterman:1977wj} and sequential recombination algorithms \cite{Catani:1993hr, Ellis:1993tq, Cacciari:2008gp, Dokshitzer:1997in, Wobisch:1998wt}. Seeded cone algorithms are fast and the resulting jets have a simple shape, which made it widely used in previous generations of colliders. The infrared/collinear (IRC) safety issues associated with the seeded algorithms can be resolved in seedless cone algorithms such as the SISCone algorithm \cite{Salam:2007xv}. Recombination algorithms, including the $k_t$ \cite{Catani:1993hr, Ellis:1993tq}, anti-$k_t$ \cite{Cacciari:2008gp}, and Cambridge/Aachen algorithms \cite{Dokshitzer:1997in, Wobisch:1998wt}, are IRC safe and have gained popularity since the fast implementations were invented \cite{Cacciari:2005hq}. A comprehensive review of various jet algorithms is given in Ref.~\cite{jetography}.  Different algorithms may yield different results, and depending on the physics goal, one jet algorithm may be more favorable than another. For example, in Ref.~\cite{Cacciari:2008gp}, it was shown that the anti-$k_t$ algorithm is less sensitive to the ``back reaction'' caused by the presence of pileup events, which results in a better measure of the jet kinematics. On the other hand, in the search for boosted Higgs, the filtering/massdrop technique \cite{BDRS} works better with the Cambridge/Aachen algorithm than other recombination jet algorithms. Therefore, there is not an algorithm that is universally advantageous. It is important to explore jet algorithms alternative to the ones commonly used in high energy experiments, which may lead to improvement in some circumstances.

Recently, Georgi proposed a class of iterative jet algorithms in Ref.~\cite{Georgi:2014zwa} for a lepton collider, which is based on maximizing a function (denoted $J$ in \cite{Georgi:2014zwa}) of the jet four-momentum. In these algorithms, all subsets of the particles in an event are considered. The subset with the maximum value of $J$ is chosen as the first jet. The particles belonging to the first jet are then removed from the event and one iterates the procedure to eventually find all jets in the event. The algorithm is ``global'' in the sense that all potential jet candidates in an event are considered simultaneously, and the one corresponding to the global maximum is chosen as the final jet. The jet algorithms have advantages to match some properties of  a QCD jet. When the $J$ function is properly chosen, the jets are infrared safe, circular and localized, making it possible to have realistic applications. Nevertheless, an implementation of the algorithms was unavailable and the $J$ function given in Ref.~\cite{Georgi:2014zwa} only works for an $e^+e^-$ machine.~\footnote{In our numerical code, we have also implemented the $J$-function-based jet-finding algorithm for a lepton collider.}  

In this paper we construct a jet-finding algorithm that can be applied at a hadron collier. Similar to Ref.~\cite{Georgi:2014zwa}, we maximize a function, $J_{E_T}=E_T- \beta\,m_J^2/E_T$, for all subsets of particles in an event. Here, $E_T$ is the transverse energy of the jet and $m_J$ is the jet mass. The motivation for this definition is similar to Ref.~\cite{Georgi:2014zwa}: we prefer a jet to have a large transverse energy but a small mass. We find that jets obtained with this definition, which we denote $J_{E_T}$ jets, have a cone shape in the Cartesian coordinates. The parameter $\beta$ controls the size of the cones. Interestingly, the axis of the cone does not coincide with the direction of the jet momentum, and shifts towards the direction perpendicular to the beam. Moreover, for a fixed $\beta$, the size of the cone shrinks when going towards the beam direction, which effectively cuts off the beam and avoids clustering excessive particles into a single jet in the forward region.

Given the simple circular shape of the jets, one can also reduce the problem of exhausting all subsets of the particles to considering all subsets that are bound by a circle on the unit sphere, which can be further simplified by considering only circles that go through three (or two) of the particles.~\footnote{This is similar to the method used in the SISCone algorithm~\cite{Salam:2007xv}, where for a fixed cone size only two particles are needed to determine a circle.} We numerically implement the jet algorithm and find it fast enough for practical applications in high energy experiments. The algorithm is infrared and collinear safe: a collinear splitting or a soft emission will not affect the function $\jet$ since it is defined as a function of the candidate jet four-momentum. The global procedure makes sure that all cones are checked to find the maximum, making it free from problems of the traditional seeded cone algorithm.   

We test the $\jet$ algorithm with several physics examples: dijets from QCD processes as well as from a heavy $Z'$ decay, and boosted all hadronic $W$ pair productions. Two different cases with and without pileup events are included in our analysis. We then compare $\jet$ to other commonly used jet algorithms, mainly the anti-$k_t$ algorithm. We find that $\jet$ and anti-$k_t$ give very similar results reconstructing the dijet momenta and the $Z'$ mass, when the jet parameter $\beta$ is chosen to match the anti-$k_t$ cone size, $R$, in the central region. As mentioned above, one of the advantages of the anti-$k_t$ algorithm is that the back reaction from soft contamination is significantly smaller than other algorithms. We find that the $\jet$ algorithm yields similarly small back-reactions as anti-$k_t$. Furthermore, we have found one advantage of $\jet$ over anti-$k_t$, which is that the $\jet$ algorithm is more efficient for capturing multiple hard components in a jet. We demonstrate this feature with the boosted $W$ example: using the same $\beta$ that is matched to the anti-$k_t$ cone size with QCD dijets, the $\jet$ algorithm has more $W$'s clustered into a single jet than anti-$k_t$.

The rest of the article is organized as follows. We give the definition of the $\jet$ algorithm and explain why the resulting jets are circular in Section \ref{sec:definition}. We describe our computing implementation in Section \ref{sec:computing-algorithm}. In Section \ref{sec:physics-example}, we test the algorithm with several physics examples, and compare it to the anti-$k_t$ algorithm. Section \ref{sec:conclusion} is devoted to the discussion of future directions and our conclusion. We have also considered other definitions of jet functions, which give us different features than $\jet$, as documented in Appendix~\ref{sec:definition-JET2}. We also show a comparison to the SISCone algorithm in Appendix~\ref{sec:siscone}.


\section{The Definition of $J_{E_T}$ for a Hadron Collider}
\label{sec:definition}
As addressed in the introduction, when combining particles to a jet, we need a function that favors a larger transverse energy or transverse momentum, but a smaller jet mass. As the simplest extension of the $J$ function for lepton colliders in Ref.~\cite{Georgi:2014zwa}, we replace energy by the transverse energy, which is more appropriate for hadron colliders, and define the $J_{E_T}$ function as
\beqa
J_{E_T} (P^\mu_J) \equiv E_T \,-\, \beta \, \frac{m_J^2}{E_T} \,.
\label{eq:JET-def}
\eeqa
Here, $P^\mu_J$ is the vector sum of the 4-momenta of the particles grouped together to form a jet candidate; $m_J$ is the jet mass; the transverse energy $E_T$ is defined as $E_T^2 \equiv P_x^2 + P_y^2 + m_J^2 = E^2 - P_z^2 = E^2 ( 1- v_J^2 \cos^2{\theta_J})$, where $v_J$ is the speed of the jet and $\theta_J$ is the angle between the jet momentum and the beam direction. The jet-finding algorithm proceeds to find the set of particles that corresponding to the global maximum of $J_{E_T}$. This set of particles is then defined as the first jet, and removed from the list of particles. We then iterate the procedure to find all jets in the event, until all particles are used. One could also try other functions with different powers of $E_T$, for instance, $J_{E_T^2}\equiv E_T\,J_{E_T}$ defined in Appendix~\ref{sec:definition-JET2}. Although we will compare the features of different definitions, we will mainly concentrate on the $J_{E_T}$ definition. The jet parameter $\beta$ is introduced to determine the geometric size of the jet and will be chosen to be $\beta \geq 1$. 

Before we discuss the detailed computing algorithm, we would like to study the general shape of the jet from our $J_{E_T}$ algorithm. Let's suppose that a group of particles with the total momentum, $P_J$, maximize the jet function $J_{E_T}$. This means that for the $j$'s particle inside this group, one has $J_{E_T}(P_J) \geq \mbox{max}[J_{E_T}(P_J - p_j), J_{E_T}(p_j)]$. To identify the geometric boundary of the jet, we consider a soft and approximately massless particle with $E_j \ll E_J$, or $r_j \equiv E_j/E_J \ll 1$. Expanding in terms of $r_j$, we have
\beqa
E_T(P_J - p_j) 
&=& E_J \sqrt{1- v_J^2 \cos^2{\theta_J}} \left[ 1 - r_j \,\frac{1-v_J\cos{\theta_J}\cos{\theta_j}}{1- v_J^2 \cos^2{\theta_J}}  \right] + {\cal O}(r_j^2) \,.
\eeqa
The condition of $J_{E_T}(P_J) \geq J_{E_T}(p_j)$ can be translated to
\beqa
1 \geq v_J^2 \geq \frac{1-1/\beta}{1- \cos^2{\theta_J}/\beta} \,.
\eeqa
This means that a central jet, with $\theta_J$ closer to $\pi/2$, can have a lower limit on $v_J$, while a forward jet is more relativistic. 

From the other condition, $J_{E_T}(P_J) \geq J_{E_T}(P_J - p_j)$, we can prove that the jet has a circular boundary on the $(\theta, \phi)$ plane.~\footnote{This is in contrast to traditional jets that are circular on the $(\eta, \phi)$ plane.} It turns out that the geometric center of the circle is not exactly along the jet momentum. Instead, it is along the direction defined by
\beqa
\vec{\hat{P}}_{c} = \frac{1}{\sqrt{1 - (1 - \kappa^2) \hat{P}^{z\, 2}_J  } } ( \hat{P}^x_J , \hat{P}^y_J , \kappa\, \hat{P}^z_J ) \,,
\eeqa
 where the unit vector is defined as $\vec{\hat{P}} \equiv \vec{P}/|\vec{P}|$. The parameter $\kappa < 1$ is a function of $v_J$ and $\theta_J$, given by
\beqa
\kappa = 1 - \frac{1}{2\beta} - \frac{1 - v_J^2}{2 - (1+\cos{2\theta_J})v_J^2} \,.
\eeqa
Denoting $z_c$ as the cosine of the angle, $\Omega_c$, between a particle $\vec{\hat{p}}_j$  and the jet geometric center $\vec{\hat{P}}_{c}$, we have $z_c$ bound by
\beqa
z_c \geq \frac{\kappa}{v_J \, \sqrt{1 - (1 - \kappa^2) \cos^2{\theta_J} }} + r_j f(v_J, \theta_J, \theta_j, \beta)\,.
\label{eq:bound-z_c}
\eeqa
One can show that the function $f$ satisfies $f(v_J, \theta_J, \theta_j, \beta) \geq 0$. So, particles belonging to the jet are confined inside a circle around the direction $\vec{\hat{P}}_{c}$. Similarly, one can prove that a particle not belonging to the jet is located outside this circle.~\footnote{The corresponding expression can be obtained by changing $\geq$ to $\leq$ and $r_j \rightarrow - r_j$ in Eq.~(\ref{eq:bound-z_c}).} Therefore, we have proved that the $J_{E_T}$ jet has a nice circular shape in Cartesian coordinates. For a central jet with $\hat{P}^z_J=0$, we have the jet geometric center matching the direction of the jet momentum. For a forward jet, on the other hand, the jet geometric center is more central compared to the jet momentum and hence stays farther away from the beam direction. This is because for particles that contribute the same amount of $E_T$ to the jet, those in the forward direction will have a larger momentum in the $z$-direction than those in the central region. Another good feature of our jet is that the cone size shrinks when the jet is closer to the beam. In the forward region with $\theta_J \rightarrow 0$, we have $z_c \equiv \cos{\Omega_c}$ with
\beqa
z_c \geq 1 - \frac{(1+1/\beta)(3-1/\beta)}{2(1-1/\beta)^2} \,\theta_J^2 \,,
\qquad
\mbox{or} \qquad 
\Omega_c < \frac{\sqrt{(1+1/\beta)(3-1/\beta) } }{1-1/\beta}\,\theta_J \,. 
\label{eq:omega-c-forward}
\eeqa
In the central region with $\theta_J = \pi/2$, the minimum value of $z_c$ can be obtained when $v_J = \sqrt{1-1/\beta}$, which is 
\beqa
z_c \geq \sqrt{1 - 1/\beta} \qquad 
\mbox{or} \qquad 
\Omega_c <  2 \arcsin{\left(\sqrt{\frac{1}{2} - \frac{\sqrt{1-1/\beta}}{2} }   \right)}
\rightarrow 1/\sqrt{\beta} \quad \mbox{for}\; \beta \gg 1   \,.
\label{eq:omega-c-central}
\eeqa

Knowing the circular shape of the $J_{E_T}$ jet, we can identify the ``fiducial" region for an individual particle to find the potential circle enclosing it. For simplicity, let us neglect the azimuthal angle and concentrate on the polar angle. There are three polar angles, $\theta_i$ for the ``$i$" particle; $\theta_J$ for the jet momentum $\vec{P}_J$ direction; $\theta_{c}$ for the geometric center direction of the jet $\hat{\vec{P}}_c$.  In terms of those three angles, we have $z_{c} \equiv \cos{(\theta_i - \theta_c) }$. The fiducial region will have an angular distance $2|\theta_i - \theta_c|$ from the ``$i$" particle. 

One of the interesting features of our $J_{E_T}$ jet is that we do not need to treat the beam direction in any special way. Although particles are copiously produced along the beam direction at hadron colliders, we can show that this is not a problem for our algorithm because the fiducial region for a particle very close to a beam will not include the beam direction. To prove that the beam will not be included is equivalent to showing $\theta_i < 2  \theta_{c}$ for the small $\theta_i$ and $\theta_{c}$ region. Noticing the relation
\beqa
\cos{\theta_{c}}  = \frac{\kappa \cos{\theta_J} } { \sqrt {1 - (1 - \kappa^2) \cos^2{\theta_J}} } \,,\qquad
\sin{\theta_{c}}  = \frac{ \sin{\theta_J} } { \sqrt {1 - (1 - \kappa^2) \cos^2{\theta_J}} } \,,
\eeqa
in the extremely forward region, we have $\theta_c \approx \theta_J/\kappa$. From Eq.~(\ref{eq:omega-c-forward}), we have
\beqa
\theta_i - \theta_c < \frac{\sqrt{(1+1/\beta)(3-1/\beta) } }{1-1/\beta}\,\kappa \, \theta_c  \approx \frac{\sqrt{(1+1/\beta)(3-1/\beta) } }{2} \, \theta_c  \,.
\label{eq:fiducial}
\eeqa
For $\beta > 1$, $\theta_i < 2\theta_c$ is satisfied, which completes our proof.  

\section{Cone-based Computing Algorithm}
\label{sec:computing-algorithm}
For a generic jet function with unknown jet shapes, one needs to check all possible subsets of the particles in an event. Each particle can be either included in or excluded from a set. Therefore for an event with $N$ particles, we easily see that there are $2^N$ such subsets to check. This becomes unrealistic once $N> {\cal O}(100)$. Fortunately, the subset of particles maximizing the $J_{E_T}$ function has been proved to be geometrically separated from the rest of the particles by a cone. This fact allows dramatic simplifications for implementing our $J_{E_T}$ jet algorithm. Instead of testing all possible combinations of the particles in an event, we only need to check those subsets that are enclosed in a cone. Thus we change the problem of finding all possible subsets to identifying all cones that contain distinct sets of particles. As we will show later in this section, we only need to consider cones uniquely determined by three particles on the boundary. It requires ${\cal O} \left( N^3 \right)$ operations to determine all possible cones, and for each cone, we also need to check whether an individual particle is inside or outside the cone. So, in total we anticipate ${\cal O} \left( N^4 \right)$ operations. In the previous section, we have shown that there exists a fiducial region for each particle. Therefore we do not need to check cones outside the fiducial region. The number of particles in the fiducial region, denoted by $n$, depends on the value of $\beta$. Then the actual number of operations is ${\cal O}\left( N n^3 \right)$. The number $n$ could be smaller than $N$ by one order of magnitude or more, which dramatically reduces the time needed by the algorithm. After calculating all $J_{E_T}$ values for all cones, one chooses the one with the largest $J_{E_T}$ as one of the final jets.  As a comparison, the number of operations needed in the $k_t$ or anti-$k_t$ algorithm is ${\cal O}(N \log N)$~\cite{Cacciari:2005hq}. Although our algorithm is substantially slower, it is still manageable, even for a single CPU with thousands of particles in an event with $\beta=6$. Our current code~\footnote{\url{https://github.com/LHCJet/JET}.} finds jets in an event containing 1000 particles with $\beta=6$ in a half second, based on a 2.6 GHz MacBook Pro.

\begin{figure}[thb!]
    \centering
    \includegraphics[width=0.8\textwidth]{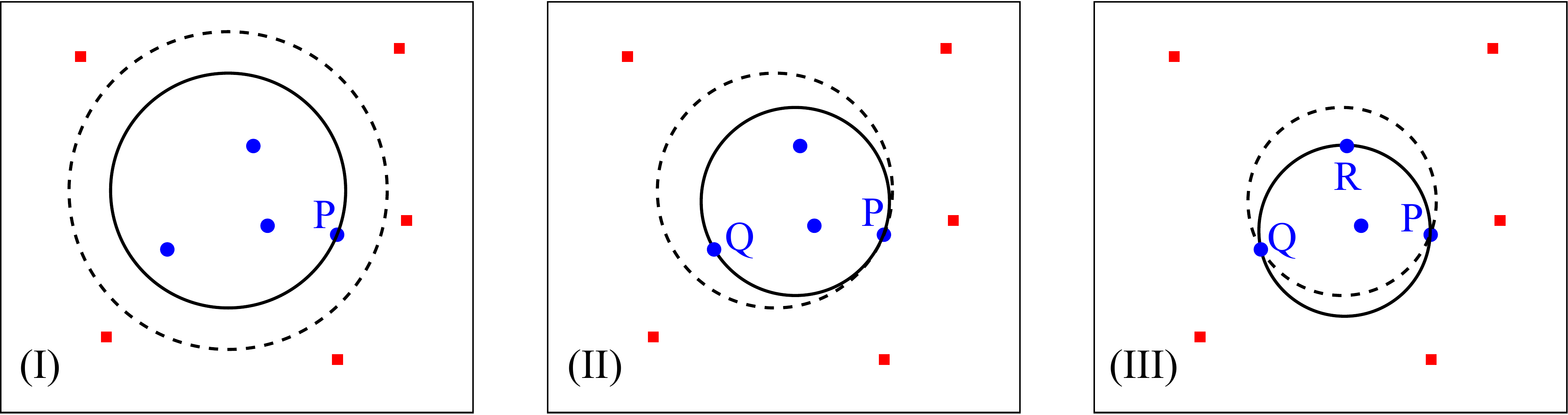}
    \caption{Specific procedure to reduce an arbitrary cone containing a subset of particles to a cone with three particles of the subset on the boundary. The blue spheres are particles inside the cone, and red squares are particles outside the cone.}
    \label{fig:proof}
\end{figure}

Before we end this section, we turn to prove that we only need to consider cones determined by three particles on the boundary. We need to show that for an arbitrary cone containing a set of particles inside, there exists another cone with three (or two) particles sitting on the boundary which contains the same set of particles (except for particles sitting on the boundary, which could be either inside or outside the original circle). This can be done by moving and deforming the original cones, as illustrated in Fig.~\ref{fig:proof}, which include the cases with three particles on the boundary. We consider a small patch in the Cartesian coordinate, and cones are represented by circles. Starting from (I) in Fig.~\ref{fig:proof}, the dashed circle is the starting cone as defined in Eq.~(\ref{eq:bound-z_c}).  The first step is to shrink the cone with the center fixed until we cross the first particle ``P". (II) Move the center of the cone towards P. In the meanwhile, shrink the cone size by keeping P on the boundary until the cone crosses the second point ``Q". (III) Move the center on the bisector of P and Q towards their midpoint, and at the same time reduce the cone size to keep P and Q on the boundary, until the cone crosses the third point ``R".  
\begin{figure}[thb!]
    \centering
    \includegraphics[width=0.8\textwidth]{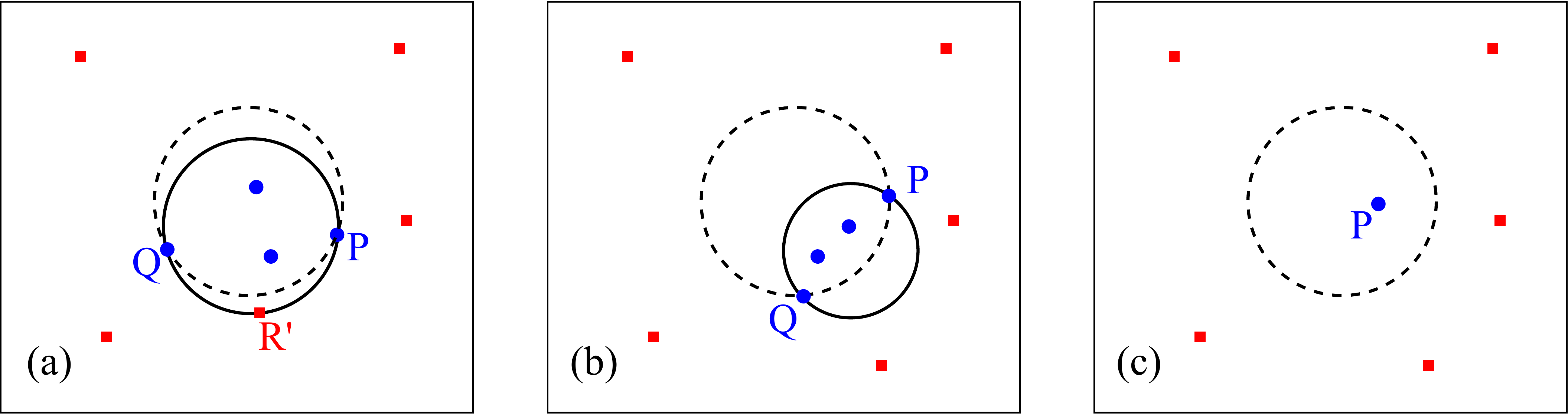}
    \caption{All alternative scenarios require special cares. Scenario (a): instead of crossing point ``R" from the insider, the cone might cross point ``R$^\prime$" from the outside first. Scenario (b): the center of the cone reaches the midpoint of ``P" and ``Q" without changing the particle content in the cone. Scenario (c): a single particle cone.}
    \label{fig:alternatives}
\end{figure}
In Fig.~\ref{fig:alternatives}, we show all alternative scenarios that require additional care. As shown in (a), in step (III), one may encounter a particle outside the original circle, denoted ``R$^\prime$'', before reaching the particle R. In (b), it happens that the circle becomes so small that the segment PQ becomes a diameter of the circle. Then one can not further reduce the cone size to reach the third particle. In the extreme case, one could have just a singe particle sitting inside a cone as depicted in (c).  

In practice our code simply finds all cones defined by three/two/one particles,  and calculates the $J_{E_T}$ function of the subset of particles that the cone encloses. Particles on the boundary require special care, since they may or may not belong to the subset. After that, one finds the global maximum of all subset $J_{E_T}$ values to determine a jet. We then remove particles associated with this jet and update all affected cones due to the removal. We then choose the cone with the next largest $J_{E_T}$ and so on, until all particles are exhausted.

\section{Physics Examples and Comparison to the anti-$k_t$ Jet}
\label{sec:physics-example}
\begin{figure}[thb!]
    \centering
    \includegraphics[width=0.6\textwidth]{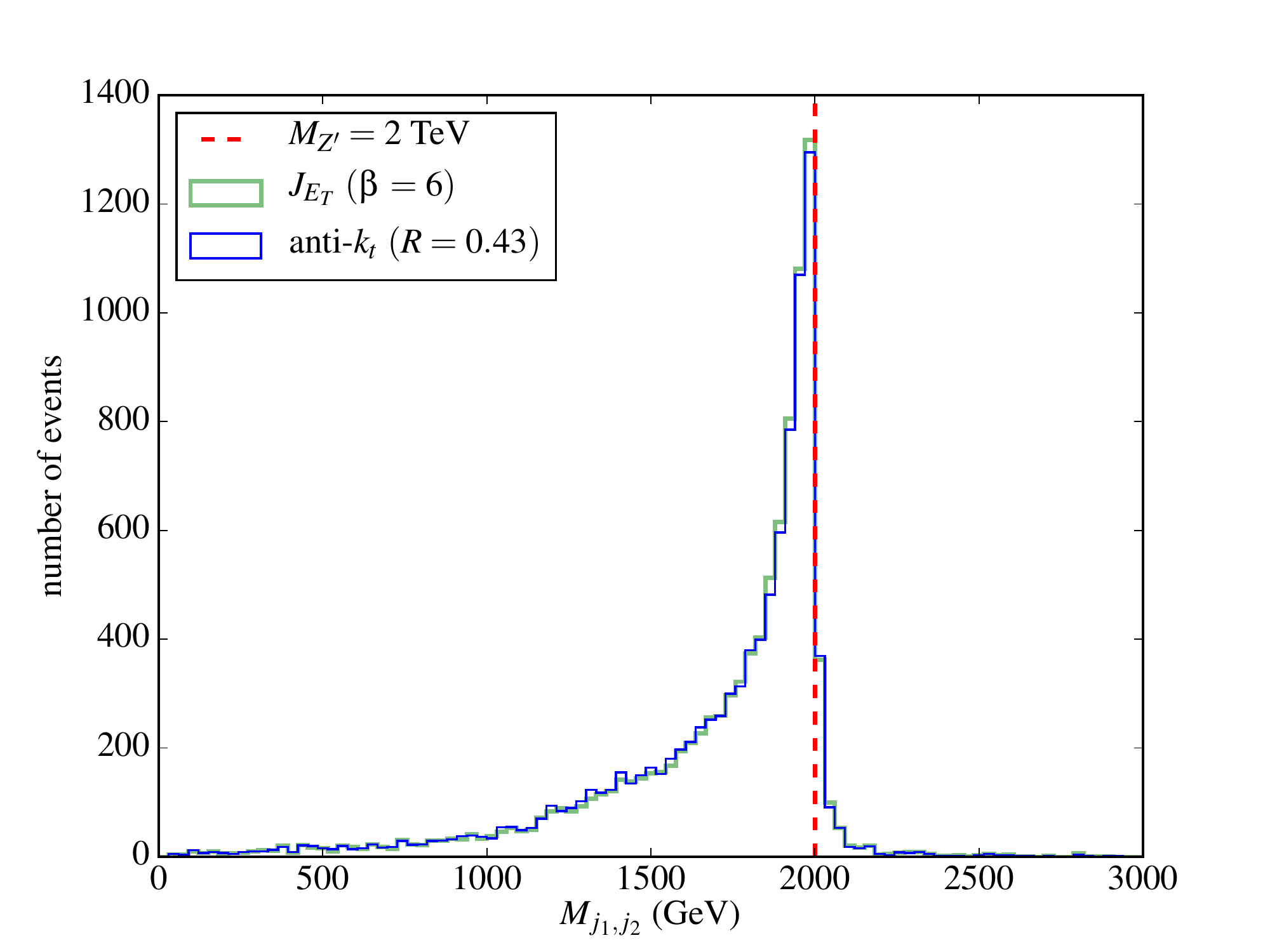}
    \caption{The invariant mass distribution for the two leading jets after the $J_{E_T}$ and the anti-$k_t$ algorithms. At parton level, a narrow $U(1)_{B-L}$ $Z^\prime$ has been simulated with $M_{Z^\prime}=2$~TeV and $\Gamma_{Z^\prime}=18$~GeV.}
    \label{fig:zprime}
\end{figure}

To demonstrate that we can use our $J_{E_T}$ algorithm to precisely reconstruct the jet momentum corresponding to a hard parton, we consider a narrow $Z^\prime$ resonance that decays to two jets. All our simulations are based on the 13 TeV LHC. We use \texttt{MadGraph}~\cite{Alwall:2011uj} to generate the parton-level dijet events from $U(1)_{B-L}$  $Z^\prime$ decays. We will compare with the anti-$k_t$ jet finding algorithm, which also gives us circular jets with the cone size determined by a parameter $R$. In Fig.~\ref{fig:zprime}, we show the reconstructed invariant mass distributions of the two leading jets for both the $J_{E_T}$ and the anti-$k_t$ jet algorithms.  For the examples in this section, we have rendered all particles massless by fixing the three-momentum of each particle and scaling its energy to match the momentum. As can be seen from Fig.~\ref{fig:zprime}, both the $J_{E_T}$ and the anti-$k_t$ jet algorithms give us a good reconstruction of the dijet $Z^\prime$ peak. The difference between the two algorithms for this distribution is barely visible.

\begin{figure}[thb!]
    \centering
    \includegraphics[width=0.6\textwidth]{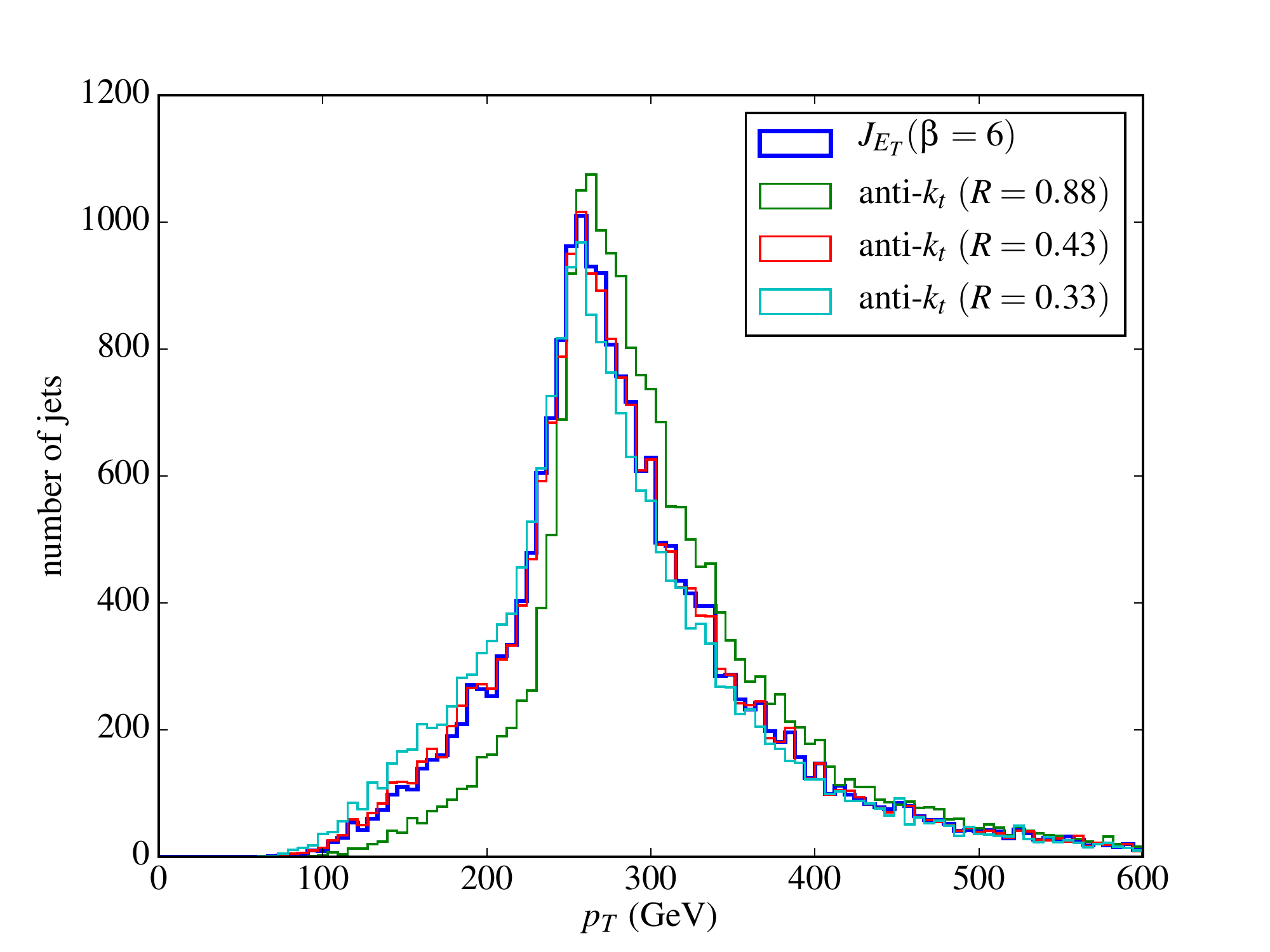}
    \caption{The leading jet $p_T$ distributions for QCD dijet events. Cuts of $p_T > 250$~GeV and $|\eta| < 5$ have been imposed on the parton-level events.}
    \label{fig:QCDdijet}
\end{figure}

In the above discussion, we have matched the parameters $\beta$ of $\jet$ and $R$ of anti-$k_t$ such that they give the same cone size in the central region. For central jets, the difference between the $\theta$ angle used in $\jet$ and the rapidity $y$ used in anti-$k_t$ becomes small. Therefore, we can use the bound on $z_c$ in Eq.~(\ref{eq:omega-c-central}) and obtain the following simple approximate relation,
\beqa
R \approx -\frac{1}{2} \log{\left( \frac{1- 1/\sqrt{\beta} }{1+ 1/\sqrt{\beta} } \right)} \,.
\eeqa
For $\beta=6$, one has $R=0.43$, which is used in Fig.~\ref{fig:zprime} and later plots. To further demonstrate the goodness of this mapping, we show the leading jet $p_T$ distributions for QCD dijet events in Fig.~\ref{fig:QCDdijet} for $J_{E_T}$ with $\beta=6$ and anti-$k_t$ with different values of $R$. To generate this plot, we have imposed cuts $p_T > 250$~GeV and $|\eta| < 5$  on the parton-level events. We see $J_{E_T}$ jets with $\beta=6$ match precisely to anti-$k_t$ jets with $R=0.43$, and the match is better than other $R$ choices.

\begin{figure}[thb!]
    \centering
    \includegraphics[width=0.45\textwidth]{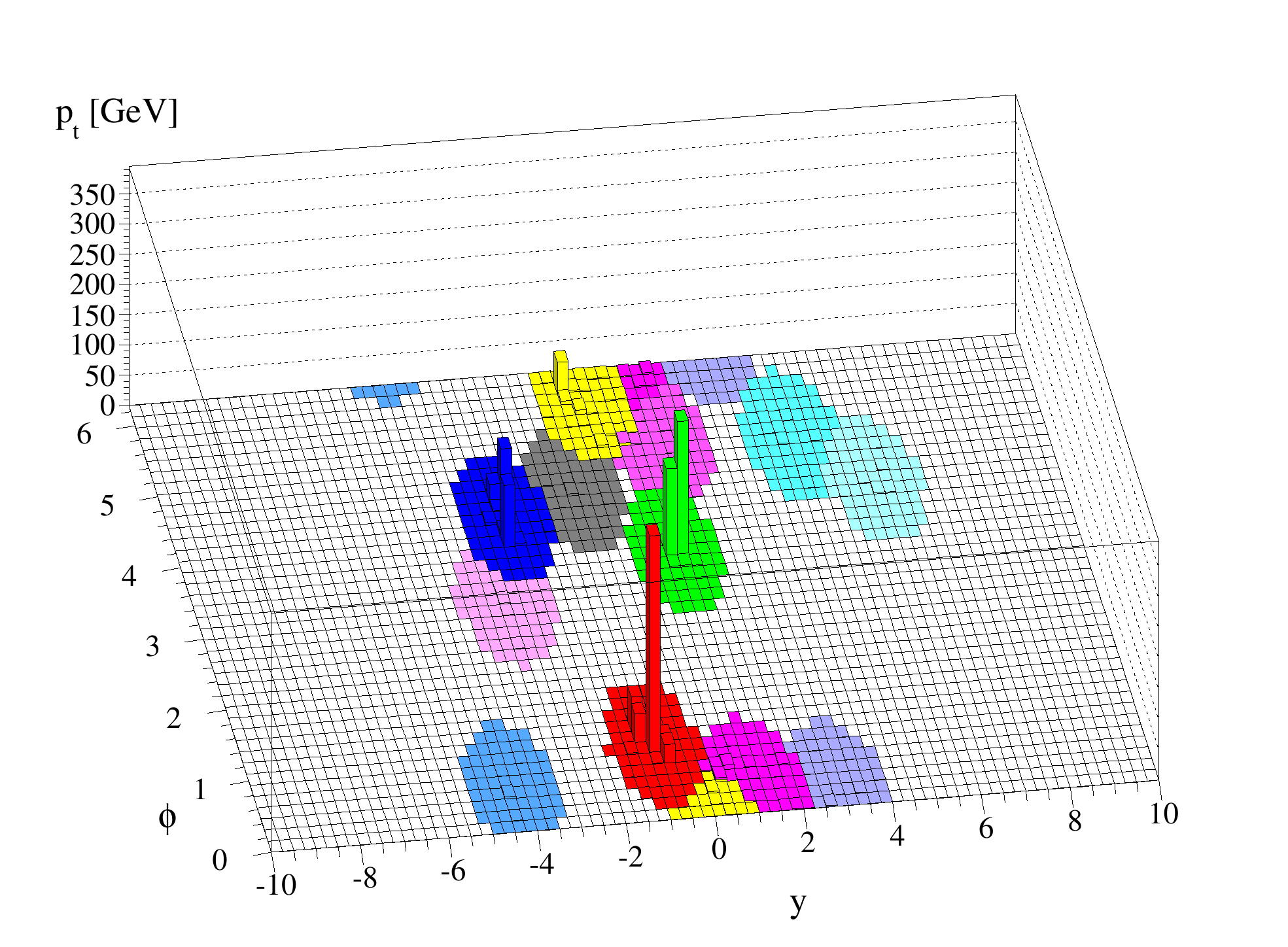}  \hspace{3mm}
        \includegraphics[width=0.45\textwidth]{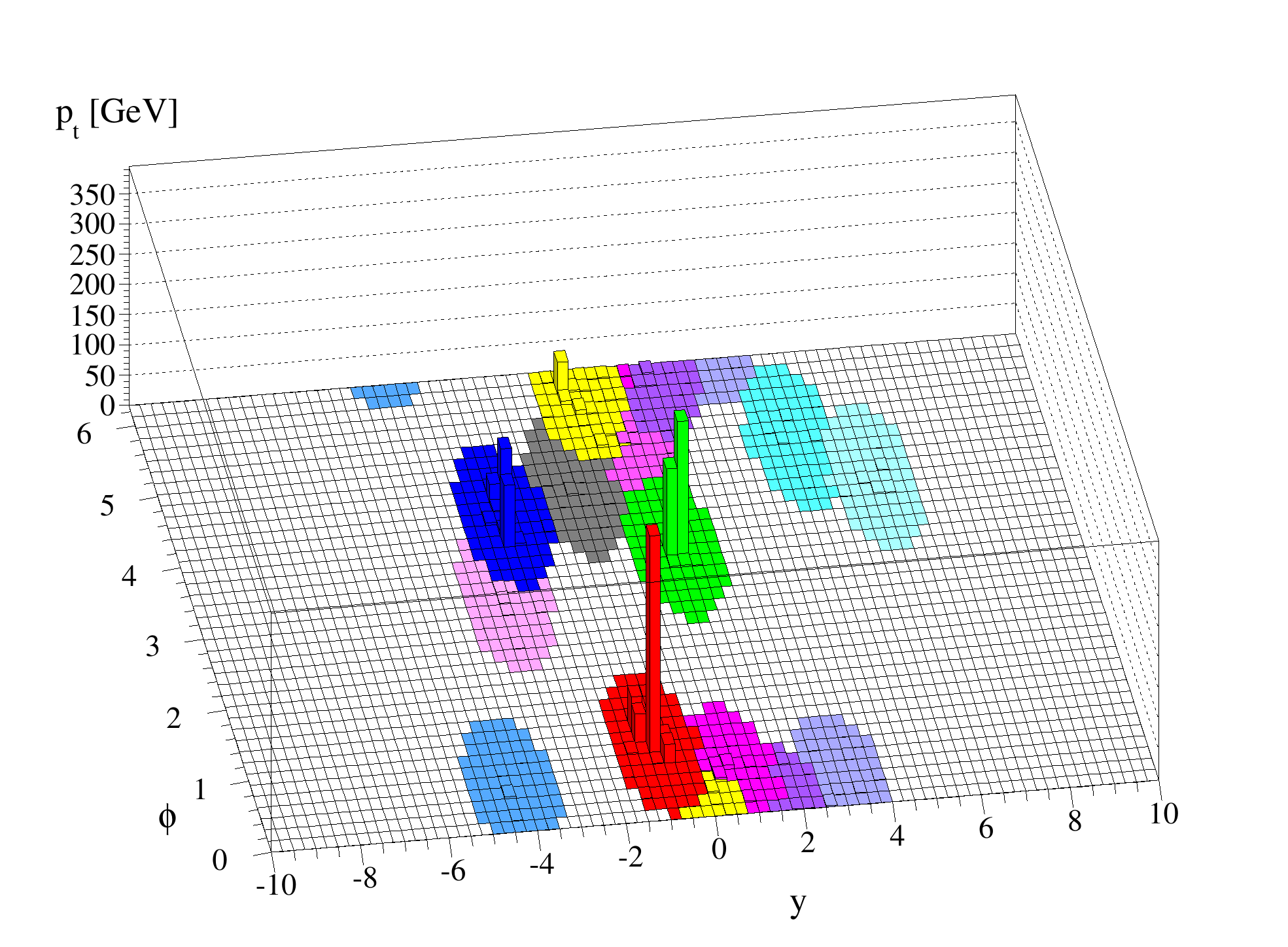}
    \caption{Left panel: an example of parton-level events with many random soft ``ghosts", grouped using $J_{E_T}$ jet algorithm with $\beta=1.4$. Right panel: the same as the left panel, but using the anti-$k_t$ algorithm with $R=1.0$.}
    \label{fig:lego}
\end{figure}

We then study the sensitivity of $\jet$ to soft contaminations such as those from underlying events and pileup, following the discussion in Ref.~\cite{Cacciari:2008gp}. We take a parton-level event with ${\cal O}(10^3)$ random soft ``ghost" particles with $p_T=1$~MeV and then show in Fig.~\ref{fig:lego} the energy deposition lego plot for the active catchment areas \cite{Cacciari:2008gn} of the resulting hard jets. Similar to the shapes of anti-$k_t$ jets, we have a regular and circular jet shape for the $\jet$ jets.  
Quantitively, one can use the so-called ``back-reaction"~\cite{Cacciari:2008gn} to quantify the modification to the hard scattering event from soft events. Specifically, one defines $\Delta p_t^{(b)}$ as the change to the summed transverse momentum of particles belonging to the hard process after including soft events. 
\begin{figure}[thb!]
    \centering
    \includegraphics[width=0.6\textwidth]{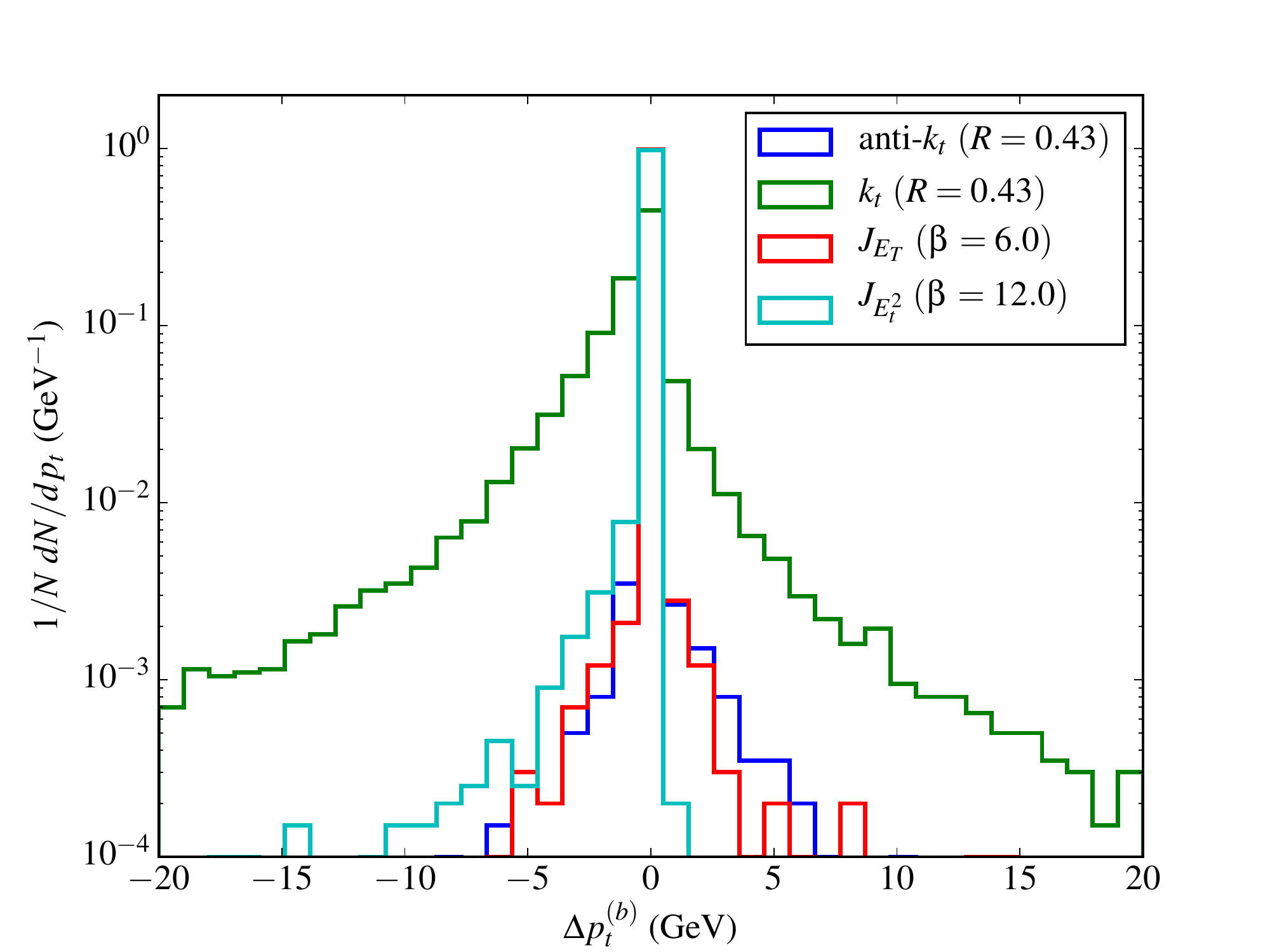}
    \caption{The normalized event distribution in terms of the back-reaction momentum $\Delta p_t^{(b)}$. It is calculated for dijet events in which the two hardest jets have $p_T > 1$~TeV and $|y| < 3$.}
    \label{fig:backreaction}
\end{figure}
In Fig.~\ref{fig:backreaction}, we show the normalized distributions of the back-reaction $\Delta p_t^{(b)}$ for different jet-finding algorithms. One can clearly see that the distribution from our $J_{E_T}$ algorithm is very close to the one from the anti-$k_t$ algorithm, both of which have a much narrower shape than the one from the $k_t$ algorithm. In Fig.~\ref{fig:backreaction}, we have also included the distribution from an alternative definition of the jet function (see the definition in Appendix~\ref{sec:definition-JET2}) which we call the $J_{E_T^2}$ algorithm. Although the distribution from the $J_{E_T^2}$ is also narrow, it is biased towards a negative value of $\Delta p_t^{(b)}$. This suggests that the $J_{E_T}$ algorithm may have some advantage over the $J_{E_T^2}$ algorithm in a busy hadron collider environment. 

\begin{figure}[thb!]
    \centering
    \includegraphics[width=0.6\textwidth]{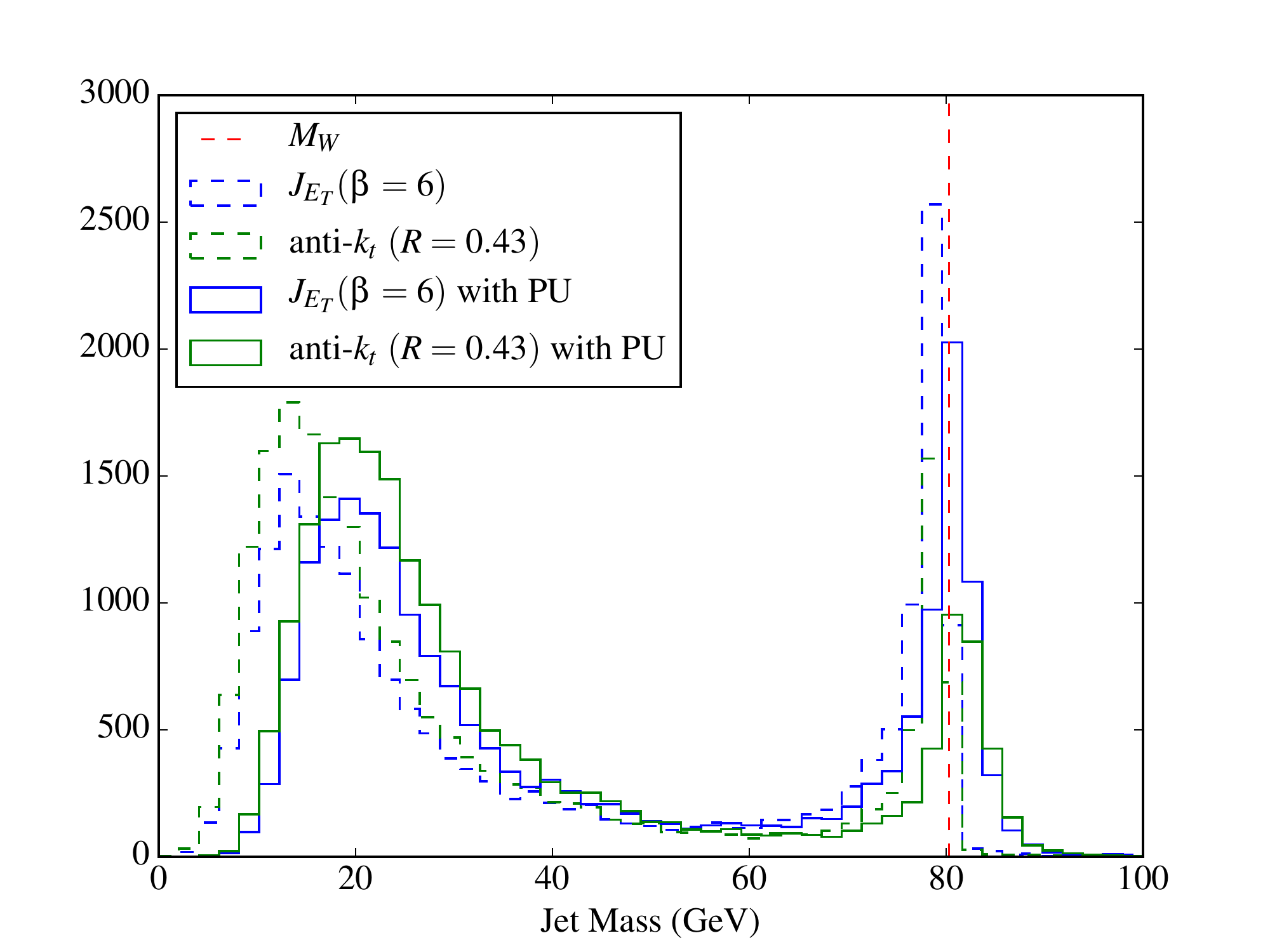}
    \caption{The event distributions in terms of jet masses for the two leading jets from $pp \rightarrow W^+ W^-$ with $p_T(W) > 250$~GeV for different jet algorithms. Solid (dashed) lines are events with (without) pile up events. The $W$ mass peaks are obtained when both partons from the $W$ decay are clustered in a single jet, while the lower peaks correspond to the case when only one hard parton is reconstructed as a single jet.}
    \label{fig:boost-w}
\end{figure}

So far, we have seen several similarities between the $J_{E_T}$ and the anti-$k_t$ algorithms. Due to its global feature, the $J_{E_T}$ algorithm also has difference from the anti-$k_t$ algorithm.  As an example, we consider boosted hadronically decaying $W$ bosons. After matching the jet cone size using QCD jets by setting $\beta=6$ for $J_{E_T}$ and $R=0.43$ for anti-$k_t$, in  Fig.~\ref{fig:boost-w} we show the jet mass distribution for the leading two jets, for boosted $W$ boson pairs with a parton level cut of $p_T(W)>250$~GeV. We have plotted the distributions both with and without pile-up (PU) events (25 minimum bias events generated with \texttt{PYTHIA6}~\cite{Sjostrand:2006za}).
From Fig.~\ref{fig:boost-w}, we see either with or without pile-up events, the $\jet$ algorithm gives us a higher $W$ mass peak than anti-$k_t$, indicating the former is more efficient for reconstructing boosted massive particles.
 
A simple way to understand the difference in Fig.~\ref{fig:boost-w} is to consider the two partons from the decay of a $W$-boson moving in the purely central direction with $p_T = |\vec{p}_W|$. The transverse energy is $E=E_T = \sqrt{p_T^2 + M_W^2}$ and the individual massless patrons have four-momenta as $p_{1, 2} = \frac{E}{2}(1, \sin{\theta}, 0, \pm \cos{\theta})$ with $\tan{\theta} = p_T/M_W$. For the $J_{E_T}$ jet, the condition to have $J_{E_T}(p^\mu_W) > \mbox{max}[J_{E_T}(p^\mu_1), J_{E_T}(p^\mu_2)]$ is $\tan{\theta} > 3.2$ or $p_T > 261$~GeV for $\beta=6$. For the anti-$k_T$ jet, we can perform a similar calculation and obtain the condition of $\tan{\theta} > 4.6$ or $p_T > 371$~GeV for $R=0.43$, which requires a more boosted $W$ to include both partons from the $W$ decay. Intuitively, this effect can be understood as follows: for two hard particles of similar $p_T$'s that are separated by a distance just above $R$, a cone with a radius $R$ and the axis sitting in between the two particles should have no problem clustering the two particles together. However, in the anti-$k_t$ algorithm, this cone configuration will be missed because the algorithm always starts from the particle with higher $p_T$ and try to include particles within a distance of $R$ around it. In the $\jet$ algorithm, because all possible cones are checked, this cone configuration will not be missed. This feature of $\jet$ may be advantageous in the search for new physics objects based on jet-substructure analysis.

\section{Discussion and Conclusions}
\label{sec:conclusion}
In this paper, we have concentrated on jet-finding algorithms at hadron colliders. The $J$-function based jet-finding algorithm in Ref.~\cite{Georgi:2014zwa} also find circular jets in Cartesian coordinates. Therefore, the three-point method to identify cones can also be applied to the lepton collider case. We have implemented this algorithm and included it in our code, which may find use at future lepton colliders such as ILC~\cite{Baer:2013cma}, FCC-ee~\cite{Gomez-Ceballos:2013zzn} and CEPC~\cite{CEPC}.

Our numerical method of using three points to identify all cones may not be the unique one. Our numerical code is practically fast enough for a jet with $\lesssim 1000$ particles, but still runs slower than commonly used jet algorithms such as SISCone and anti-$k_t$.  Therefore, it would be very useful if a faster implementation could be found. Some techniques using local information of particles, such as those utilizing the Voronoi diagram to identify the nearest neighbors \cite{Cacciari:2005hq}, do not work for our jet algorithm due to the global nature of our jet definition. 

One of the obvious follow-ups is to identify more features about this global $J_{E_T}$ algorithm, and find its possible advantages for tagging boosted objects such as the $W$-boson, the Higgs boson and the top quark. Another direction could be to modify the $J_{E_T}$ definition to include the masses of boosted objects and apply different functions to identify different objects. The subset maximizing a generic $J$ function may not be a cone, but we can always restrict to cones such that the algorithm described in this article can be easily adopted for a different $J$ function. We leave them as future studies. 

In summary, we have studied a new jet-finding algorithm based on the $J_{E_T}$ function that prefers a larger value of the jet transverse energy and a smaller value of the jet mass. Our algorithm is a global one in the sense that one needs to compare all possible subsets of particles to find the global maximum value of $J_{E_T}$. We have found that our $J_{E_T}$ jet shape is a circle with the geometric center shifted in the central direction from the jet momentum. The jet cone size also shrinks when it is closer to the beam direction. No special attention is needed for the beam direction, because the fiducial region of an individual particle can be shown to exclude the beam direction. We use the three-point way to identify all cones to make our algorithm practically useful. The general running time is ${\cal O}(N n^3)$ with ``$N$" as the total number of particles and ``$n$" as the number of particles in the fiducial region of an individual particle. For QCD jets, we have found that our $J_{E_T}$ jet is similar to the anti-$k_t$ jet for many features including the ``back-reaction" event distribution. For a boosted $W$-jet and compared to the anti-$k_t$ jet, our $J_{E_T}$ jet algorithm could tag a boosted massive jet in a more efficient way.

\subsection*{Acknowledgments}
We would like to thank Vernon Barger and Gavin Salam for useful discussion and comments. Y.~Bai and R.~Lu are supported by the U. S. Department of Energy under the contract DE-FG-02-95ER40896. Z.~Han is in part supported by U.~S.~Department of Energy under grant numbers DE-FG02-96ER40969 and DE-FG02-13ER41986.

\appendix
\section{The Alternative $J_{E_T^\alpha}$ Function}
\label{sec:definition-JET2}
The jet function defined in Eq.~(\ref{eq:JET-def}) can be extended to other forms. For example, one could define the $J$ function as $E_T^\alpha - \beta m_J^2/E_T^{2-\alpha}$ to weight the jet mass in different ways (see Ref.~\cite{Ge:2014ova} for other definitions). For $\alpha=1$, we recover our $J_{E_T}$ definition. For a general $\alpha$ with $0 \leq  \alpha \leq 2$, we can prove that particles included in the jet are within a cone and particles that does belong to the jet are outside the cone. The geometric center is 
\beqa
\vec{\hat{P}}_{c} = \frac{1}{\sqrt{1 - (1 - \kappa^2) \hat{P}^{z\, 2}_J  } } ( \hat{P}^x_J , \hat{P}^y_J , \kappa \hat{P}^z_J ) \,.
\eeqa
with
\beqa
\kappa = 1 - \frac{\alpha}{2\beta} + \frac{\alpha-2}{2} \,\frac{m_J^2}{E_T^2(P_J)} \,.
\eeqa
The condition of $J_{E_T^\alpha}(P_J) \geq J_{E_T^\alpha}(p_j)$ provides the constraint
\beqa
1 \geq v_J^2 \geq \frac{1-1/\beta}{1- \cos^2{\theta_J}/\beta} \,.
\eeqa
From the other condition $J_{E_T^\alpha}(P_J) \geq J_{E_T^\alpha}(P_J - p_j)$, we obtain a constraint on the cosine of the angular distance of a particle $\vec{\hat{p}}_j$ to the jet geometric center $\vec{\hat{P}}_{c}$ as
\beqa
z_{c}   \geq   \frac{\kappa}{v_J\,\sqrt{1 - (1 - \kappa^2) \cos^2{\theta_J}  } } \,.
\eeqa

For the special case with $\alpha=2$ and in the forward region with $\theta_J \rightarrow 0$, we have 
\beqa
z_c \geq \frac{1}{v_J} \left[ 1 - \frac{2\beta-1}{2(\beta-1)^2}\, \theta_J^2 \right] \geq 1 - \frac{2\beta-1}{2(\beta-1)^2}\, \theta_J^2
\, \qquad
\mbox{or} \qquad
\Omega_c  \leq \frac{\sqrt{2\beta -1}}{\beta -1 } \theta_J  \,.
\eeqa
This shows that the jet cone size shrinks as it becomes more forward. In the central region, the constraint becomes
\beqa
z_c \geq \frac{1}{v_J} \left( 1 - 1/\beta \right)  \geq  1 - 1/\beta  
\, \qquad
\mbox{or} \qquad
\Omega_c  \leq 2 \arcsin{ \left(1/\sqrt{2\beta}\right)  }  \rightarrow \sqrt{2/\beta}\quad \mbox{for}\;\beta \gg 1\,.
\eeqa
Comparing to Eq.~(\ref{eq:omega-c-central}) for the $J_{E_T}$ case in the large $\beta$ limit, we need to choose a $\beta$ value for $J_{E_T^2}$ twice as big as that for the $J_{E_T}$ case to have the same cone size. Finally, for the fiducial region of a particle to stay away from the beam direction, one needs to impose $\beta  > 2$ for the $J_{E_T^2}$ algorithm.

\section{A Comparison to the SISCone Algorithm}
\label{sec:siscone}
Since our algorithm is based on searching for all possible cones, it bares some similarities to the SISCone algorithm~\cite{Salam:2007xv}, which we compare to in this appendix. The SISCone algorithm works by first finding all ``stables'' cones with a fixed cone size, that is, cones with the axis coincident with the direction of the sum of the 4-momentum of the particles it encloses. Then the algorithm uses a Tevatron Run-II type split-merge procedure \cite{Blazey:2000qt} to treat overlapping stable cones. Roughly speaking, a parameter $f$ controls the probability whether one wants to split or merge two overlapping cones. A larger $f$ means a smaller probability to merge, and when $f=1$, overlapping cones are always split.  

\begin{figure}[thb!]
    \centering
    \includegraphics[width=0.6\textwidth]{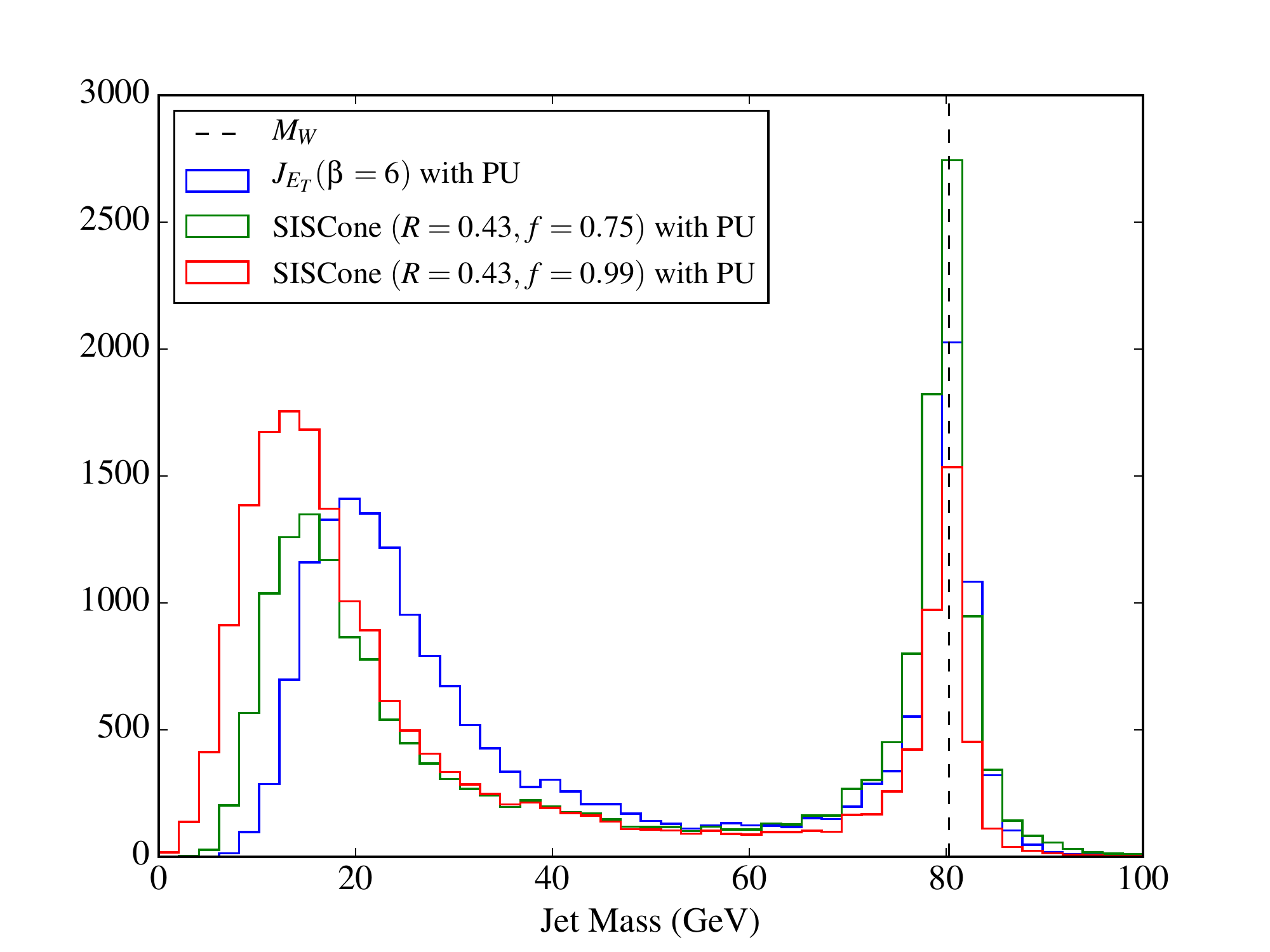}
    \caption{The leading jet mass for the two leading jets from $pp \rightarrow W^+ W^-$ with $p_T(W) > 250$~GeV for the $\jet$ algorithm and the SISCone algorithm with  $f=0.75$ and $f=0.99$, with 25 pileup events per hard event.}
    \label{fig:boost-w-siscone}
\end{figure}

Using QCD dijet events with a $p_T>250~\gev$ parton level cut, we have found that SISCone jets with a cone size $R=0.43$ match to our $J_{E_T}$ jets with $\beta=6$ very well. The two algorithms give almost indistinguishable $p_T$ distributions. For the case of a boosted $W$, the definition of ``stable'' cones for SISCone without the split-merge procedure do not prefer to include a $W$ in a single cone. However, the split-merge procedure of SISCone allows the possibility for two cones to merge to form a bigger jet with a radius larger than $R$. Therefore, for a smaller $f$, we have more $W$'s clustered in a single jet. This effect is shown in Fig.~\ref{fig:boost-w-siscone}, where we compare our $\jet$ algorithm with SISCone for two different $f$ values, $f=0.99$ and $f=0.75$, with pileup. We see that the efficiency for finding a $W$ in a single jet using the $\jet$ algorithm is smaller than SISCone with $f=0.75$, while larger than $f=0.99$ where overlapping cones are almost never merged. Since in our $\jet$ algorithm the cone with the largest $\jet$ is always taken as a separate jet, we do not have the freedom to obtain a jet of larger size. Therefore, it is more appropriate to compare with the $f=0.99$ case of SISCone. If our goal is to increase the $W$-jet identification efficiency, a split-merge procedure can also be adopted in the $\jet$ algorithm. It is interesting to conduct a more detailed study to determine which algorithm is more favorable for $W$-jet tagging.

\bibliography{JtJET}
\bibliographystyle{JHEP}

 \end{document}